\documentclass[pra,amsmath,twocolumn,showpacs]{revtex4-1}

\usepackage{bm}
\usepackage{amssymb}
\usepackage{graphicx,xcolor}
\usepackage{amsmath,graphicx}
\begin{document}
\def\la{{\langle}}
\def\U{\hat U}
\def\lm{\lambda}
\def\q{\quad}
\def\t{\tau}
\def\n{\\ \nonumber}
\def\ra{{\rangle}}
\def\Ep{{\mathcal{E}}}
\def\omga{{\epsilon}}
\def\t{{\tau}}
\def\h{\hat{H}}
\title{Paths, negative "probabilities", and the Leggett-Garg inequalities}
%
%
\author {D. Sokolovski$^{a,b}$} 
\email {dgsokol15@gmail.com}
\author {S.A. Gurvitz$^c$}
\affiliation{$^a$ Departmento de Qu\'imica-F\'isica, Universidad del Pa\' is Vasco, UPV/EHU, Leioa, Spain}
\affiliation{$^b$ IKERBASQUE, Basque Foundation for Science, E-48011 Bilbao, Spain}
\affiliation{$^c$ 
Department of Particle Physics and Astrophysics, Weizmann Institute of Science, Rehovot, 76100, Israel}
\begin{abstract}
\noindent
{We present a path analysis of the condition under which the outcomes of previous 
observation affect the results of the measurements yet to be made.
It is shown that this effect, also known as "signalling in time", occurs whenever the earlier measurements 
are set to destroy interference between two or more virtual paths. 
We also demonstrate that Feynman's negative "probabilities" 
provide for a more reliable witness of "signalling in time", than 
 the Leggett-Garg inequalities, while both methods are frequently subject to failure.}
 \date\today
\end{abstract}

%
%
\maketitle
%
%
\vspace{0.1cm}
Recently, the authors of \cite{nat} have shown that superconducting flux qubits possess,
despite their macroscopic nature, such quantum properties, 
as the ability to exist in a superposition of distinct states.
After reviewing an approach based on the so-called Leggett-Garg inequalities (LGI),
which may or may not be satisfied by certain quantum mechanical averages \cite{LGI}, 
they chose to employ a simpler experimental protocol. The method used in \cite{nat}
was similar to the one proposed by  Koffler an Bruckner   \cite{Koff} , who suggested that the 
relevant evidence can be obtained more efficiently by analysing corresponding probability 
distributions, and coined a term "signalling in time".
\newline 
Both the LGI, and the notion of "signalling", are closely related to a different 
problem, the so-called Bell test \cite{Bell}, in which Alice an Bob are given two 
spins in the zero total spin state. Alice's measurement along a chosen axis
immediately aligns Bob's spin in the opposite direction, and the study of 
inequalities, formally similar to the LGI, allowed Bell to show that the phenomenon
cannot be explained by the existence of certain classical-like hidden variables.
There is, however, no "signalling" in the Bell's experiment, in the sense 
that Bob is unable to recognise the choice of the axis made by Alice.
Reduced density matrix of the Bob's spin is not affected by Alice's decision,
and the no-cloning theorem \cite{noclo}, constraints his ability to 
reconstruct the spin's state.
\newline
Feynman's approach to Bell's problem  has been more direct.
In the \cite{Feyn} he demonstrated that, in order to reproduce quantum 
results for an entangled Bell's state with hidden variables, some the probabilities 
would inevitably turn negative. In a recent essay on the relationship between 
Feynman and Bell \cite{Whit}, Whitaker notes that "what Feynman describes is
indeed Bell's Theorem". {\color{black} A similar, yet somewhat different approach to the "signalling in time"
problem was recently proposed in \cite{Hall}, where negative values taken by quasi probabilities,
defined in terms of quantum projection operators, were related
with violations of the LGI. }
\newline
Several authors \cite{Koff},\cite{Hall},\cite{Em1}, 
emphasise the difference between the Bell's case, and the problem, to which the LGI is usually applied.
Indeed, here one makes several consecutive measurements on the same quantum system, 
and asks whether the outcomes of previous observations can influence the results of the measurements
yet to be made. Interaction with a measurement device at some $t_1$ can scatter the system, at $t_2>t_1$
 into a state it would not have visited otherwise, or visit it with a different frequency. 
 Should this happen, "signalling in time" is said to have occurred \cite{Foot}. 
 No "signalling" means  that a measurement does not change the outcome statistics 
 of later measurements \cite{Koff}.
 \newline
 The literature on the Leggett-Garg inequalities is extensive, and we refer the reader 
to a recent review \cite{Em2} for relevant references, covering different aspects of the problem.
The scope of this paper is much narrower. First, we analyse "signalling in time" in terms 
of the virtual (Feynman) paths, and illustrate the analysis on the simple example of a qubit
undergoing Rabi oscillations.
 Having done so, we compare the Feynman's direct "negative probability" test, and the violation of the LGI, as possible indicators of the "signalling" phenomenon.
\section {Path analysis of "signalling in time".}
Consider a sequence of accurate measurements of quantities \{$\hat Q_1$,$\hat Q_2$, ...,$\hat Q_K$ \}
which could, in principle made on a quantum system in a Hilbert space of a dimension $N$ at
different times, \{$t_1$,$t_2$, ...,$t_K$ \}.
Let us call a {path} a sequence of possible measurement outcomes (numbers), 
\{$Q_1$,$Q_2$, ...,$Q_K$ \}, where $Q_i$, $1\le i\le K$ is one of the eigenvalues of the operator $\hat Q_i$.
(The simplest paths would connect just two outcomes, e.g., $Q_2 \gets Q_1$).
 For every path quantum mechanics provides a complex valued {\it probability amplitude}, 
  $A(Q_1,Q_2, ...,Q_K)$.
If all  (or possibly some) of the measurements are actually made, 
quantum mechanics provides also
the probabilities  \cite{Feynl}, e.g., 
$P(Q_1,Q_2, ...,Q_K)=|A(Q_1,Q_2, ...,Q_K|^2$,
if all of the 
measurements are realised.
The probabilities are related to the frequencies, 
with which a given sequence will be observed, and the system will be seen to "travel" the corresponding path.  
While paths endowed only with probability amplitudes are usually called {\it virtual}, it seems 
reasonable to describe the paths, to which both the amplitude and the probability, as {\it real} \cite{DS1}.
We will call the set of all relevant real paths, together with the corresponding probabilities, a {\it statistical ensemble}.
The probabilities for real paths can be obtained by adding the probability amplitudes of the virtual paths,
and taking the absolute square, as appropriate \cite{Feynl}. Importantly, choosing to make different measurements 
from the set \{$\hat Q_1$,$\hat Q_2$, ...,$\hat Q_K$ \} may lead to essentially different statistical ensembles \cite{DS1}, \cite{DS2}. Now the problem of "signalling in time" can be seen as follows. Two measurements 
at $t_1< t_3$ produce and ensemble with $N^2$ real paths. Adding a third measurement at $t_1<t_2<t_3$
yields another ensemble, with $N^3$ real paths. This ensemble is different, i.e. incompatible, with the first one, in the sense that  ignoring the outcomes at $t_2$ (non-selective measurement), and adding the corresponding probabilities, does not recover the ensemble, obtained with the measurements made at $t_1$ and $t_3$ only. Incompatibility of different ensembles comes as a natural consequence of the fact that one inevitably perturbs an accurately measured quantum system. 
\section{Virtual and real paths for a qubit}
As an example, consider a two-level system, such as spin-$1/2$ (a qubit), and three consecutive measurements,
made at $t=0$, $t=\tau$, and $t=T$,
of the same quantity $\hat Q=|+\ra\la+|-|-\ra\la-|$, which can take the values of $\pm 1$.
For simplicity, we will assume that at $t=0$ the system is in an eigenstate of $\hat Q$, 
so that the result of the first measurement is always $Q_1=+1$. Now there are four virtual paths
shown in Fig.1, 
\begin{eqnarray} \label{0.1}
\{1\} \equiv \{\q 1,\q 1,\q1\},\n
\{2\} \equiv \{\q 1,-1,\q1\},\n
\{3\} \equiv \{-1,\q1,\q 1\},\n
\{4\} \equiv \{-1,-1,\q 1\},\n
\end{eqnarray}
endowed with the probability amplitudes
\begin{eqnarray} \label{0.2}
A[1] \equiv \la +|\U_{T-\t}|+\ra\la+|\U_\t|+\ra=\cos(T-\t)\cos\t,\q\n
A[2] \equiv \la +|\U_{T-\t}|-\ra\la-|\U_\t|+\ra=-\sin(T-\t)\sin\t,\n
A[3] \equiv \la -|\U_{T-\t}|+\ra\la+|\U_\t|+\ra=-i\sin(T-\t)\cos\t,\n
A[4] \equiv \la -|\U_{T-\t}|-\ra\la-|\U_\t|+\ra=-i\cos(T-\t)\sin\t,\n
\end{eqnarray}
Here we have let the system performs Rabi oscillations of a unit frequency, $\omega=1$, 
between the states $|+\ra$ and $|-\ra$, so that its evolution operator in Eq.(\ref{0.2}) can be written as
\begin{eqnarray} \label{0.3}
\U(t)=  \cos t \hat I -i\sin t \hat \sigma_x, 
\end{eqnarray}
with $\hat I$ and $ \hat \sigma_x$ denoting the unity, and Pauli $x$-matrix, respectively.
\newline 
We will consider three sets of measurements, 
(i) made at $t_1=0$ and $t_3=\tau$, thus yielding the values $Q_1$ and $Q_2$, 
\newline
(ii) made at $t_1=0$ and $t_3=T$, 
yielding $Q_1$ and $Q_3$, and 
\newline
 (iii) made at $t_1=0$,  $t_2=\t$ and $t_3T$, 
yielding $Q_1$, $Q_2$ and $Q_3$. 
The corresponding statistical ensembles are shown in Fig.2. 
In the case (ii)  there are just two real paths, $\{I\}=\{1,1\}$ and $\{ II \}=\{-1,1\}$, given by the superpositions of the paths
of $\{1\}$ and $\{2\}$, and of $\{3\}$ and $\{4\}$, respectively (see Fig. 2b).  The corresponding probabilities, therefore,  are
\begin{eqnarray} \label{0.4}
P[I] = |A[1]+A[2]|^2=\cos^2(T),\n
P[II] = |A[3]+A[4]|^2=\sin^2(T).
\end{eqnarray}
In the case (iii), all four paths in Eqs.(\ref{0.1}) become real (see Fig. 2c), and are travelled with the probabilities
\begin{eqnarray} \label{0.5}
P[1] =\cos^2(T-\t)\cos^2(\t),\n
P[2] =\sin^2(T-\t)\sin^2(\t),\n
P[3] =\sin^2(T-\t)\cos^2(\t),\n
P[4] =\cos^2(T-\t)\sin^2(\t).
\end{eqnarray}
Finally, since future measurements do not affect current results, the probabilities 
of the two real paths in the case (i),  $\{I\}$ and  $\{II\}$ in Fig. 2a, can be found 
by summing over the outcomes at $t=T$, shown in Fig. 2c, 
\begin{eqnarray} \label{0.6}
P'[I] = P[1]+P[3]=\cos^2(\t),\n
P'[II] =P[2]+P[4]=\sin^2(\t).
\end{eqnarray}
\begin{figure}[h]
\includegraphics[angle=0,width=8cm, height= 6cm]{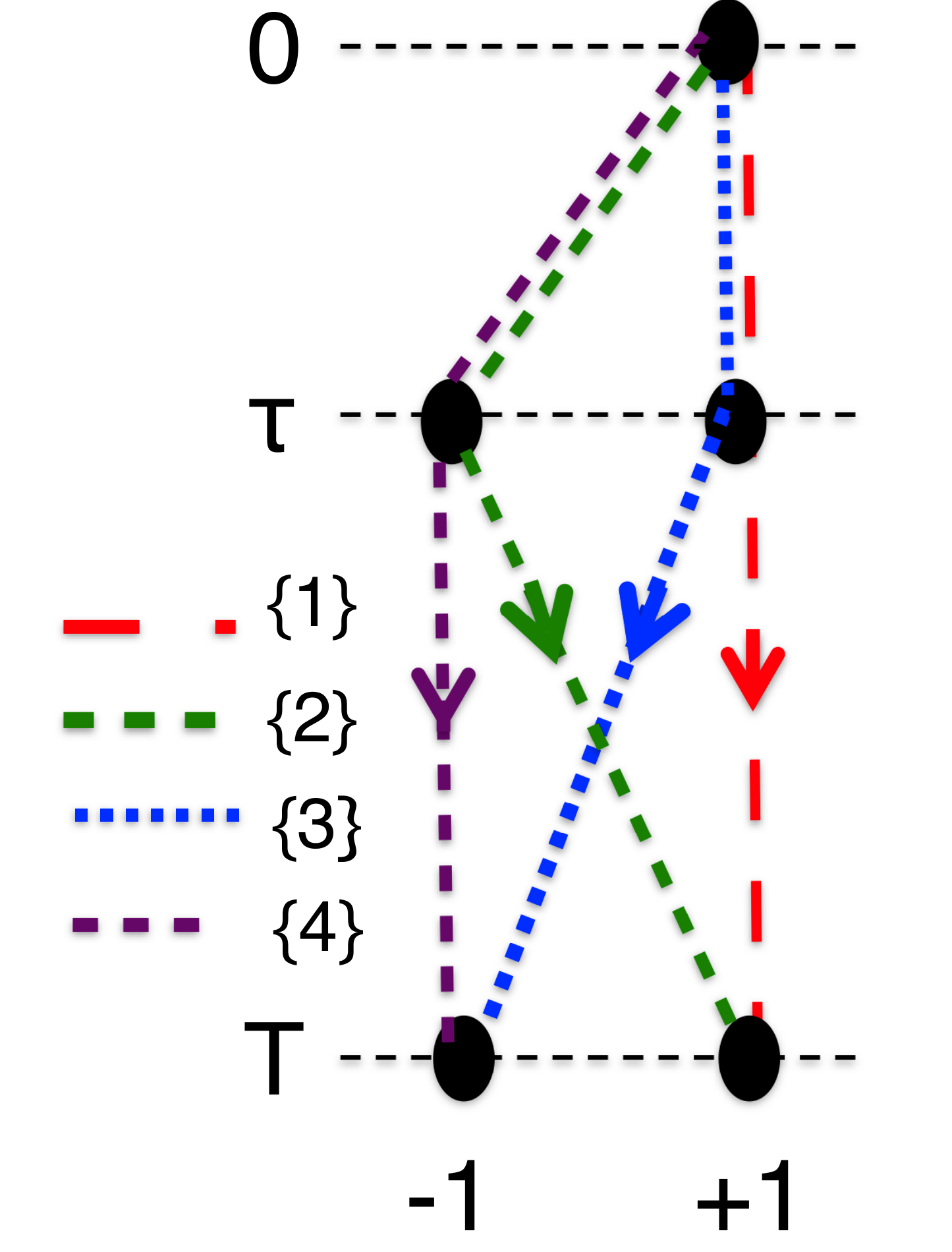}
\caption {Four virtual paths in Eq.(\ref{0.1}), for the chosen sets of measurements.}
\label{fig:FIG1}
\end{figure}
\begin{figure}[h]
\includegraphics[angle=0,width=8.5cm, height= 4.5cm]{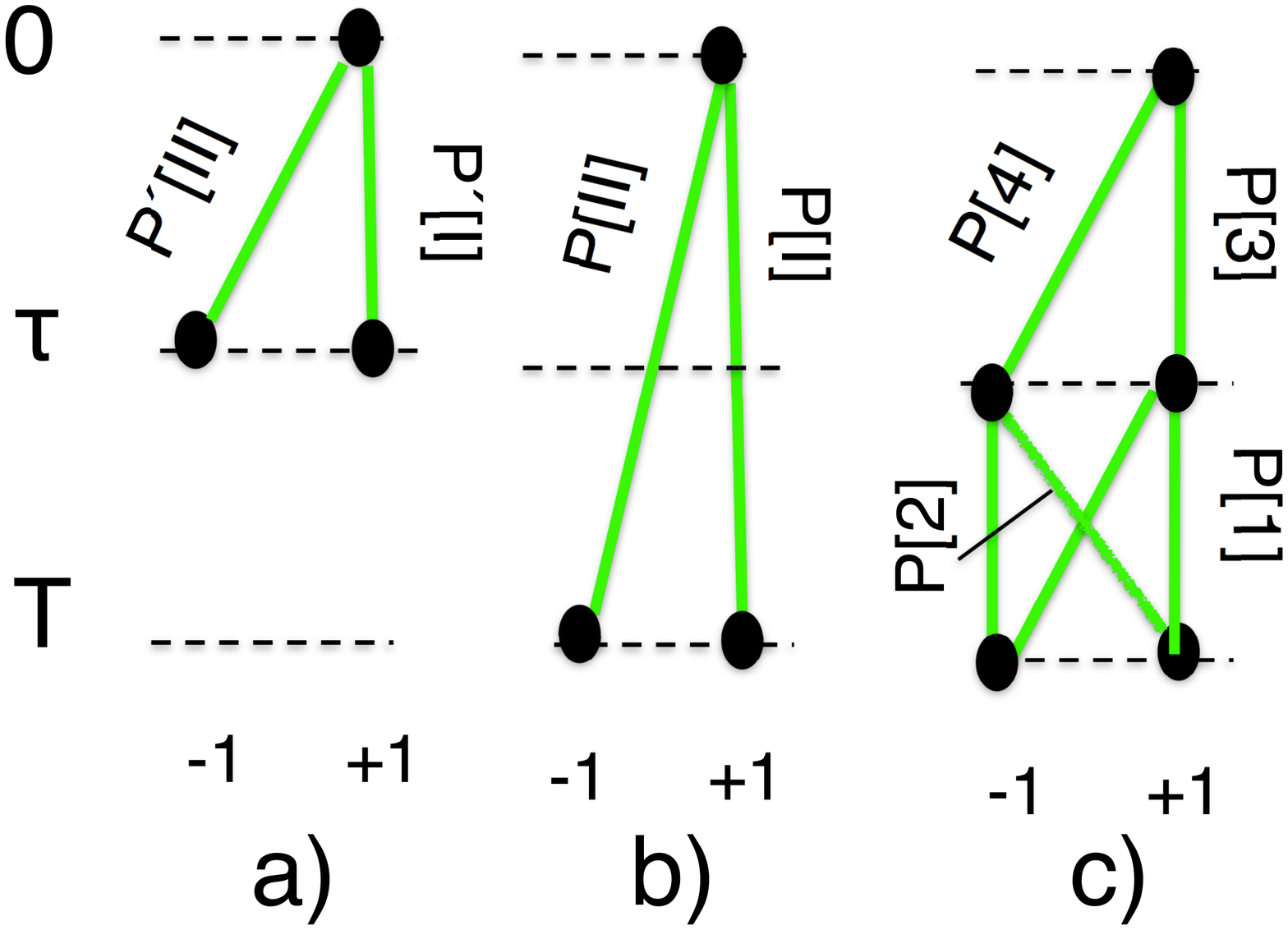}
\caption {Real paths in Eqs.(\ref{0.4})-(\ref{0.6}).}
\label{fig:FIG1}
\end{figure}
\begin{figure}[h]
\includegraphics[angle=0,width=8.5cm, height= 7cm]{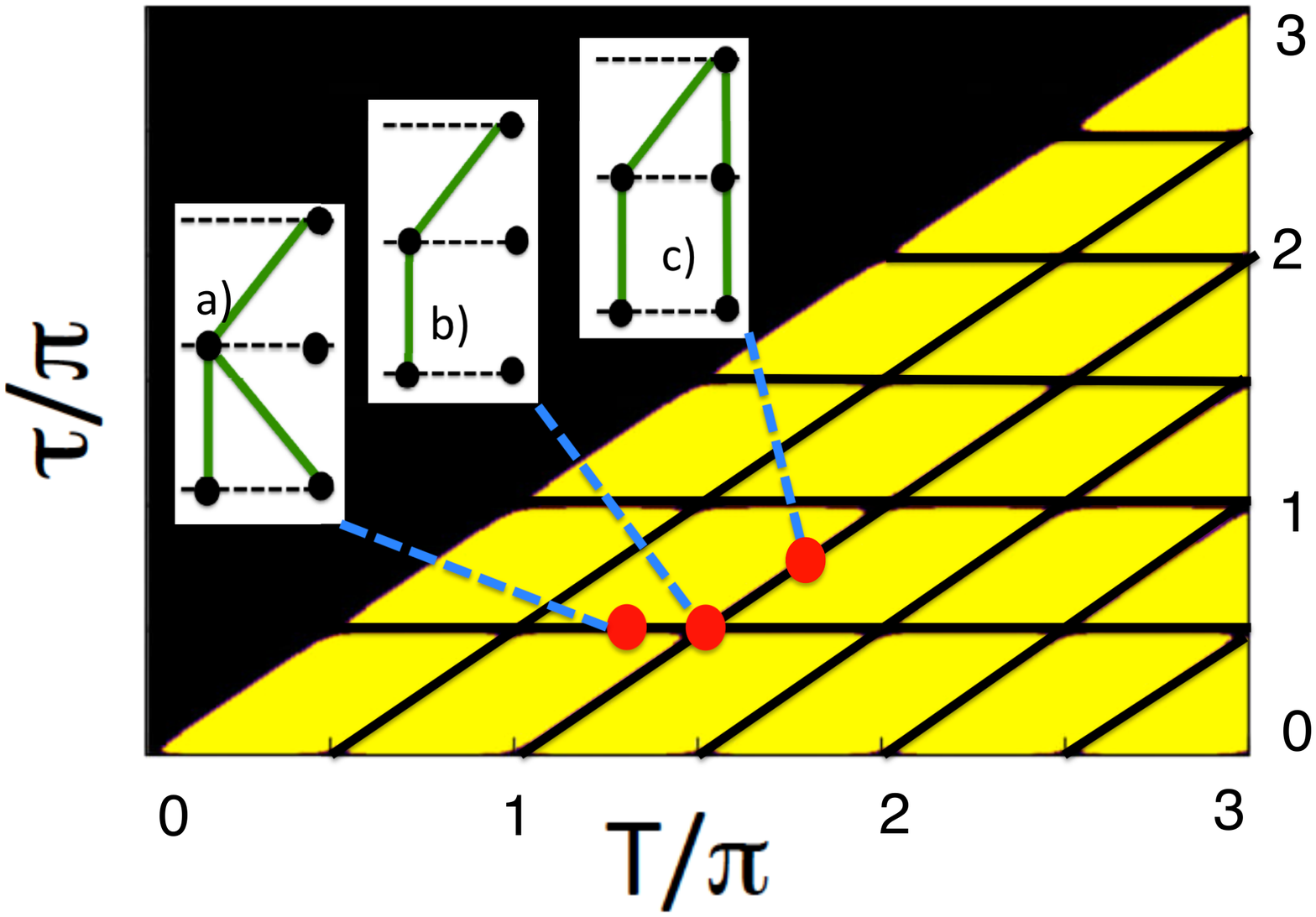}
\caption {Detection of "signalling in time"  based on Eq.(\ref{0.7}).
In the light coloured area $\delta P(Q_3=1\ne 0$, and
the ensemble consists of the four real paths shown in Fig.2c. 
On the dark coloured horizontal and diagonal lines, the number of real paths
is reduced to two [see insets a) and c)]. At their intersection, 
there is only one real path, as shown in the inset b).}
\label{fig:FIG1}
\end{figure}
\newline
Our aim
is to identify the conditions under which 
 the ensembles in Figs. 2 are found  to be
essentially different.
In other words, we want to know \cite{LGI}-\cite{Koff},   \cite{Hall}-\cite{Em1},  \cite{Em2} when
making the measurement of 
$Q_2$ at $t=\t$ affects the distribution of the values $ Q_3$ at $t=T$. 
It is sufficient to look at the probability to have $Q_3=1$ in the case c) shown in Fig.2,
\begin{eqnarray} \label{0.7}
Prob(Q_3=1)= P[1]+P[2]=
 cos^2(T)+\frac{1}{2}\sin(2\t)\sin(2(T-\t))\n
 Prob_0 (Q_3=1)+\delta P(Q_3=1).\q\q\q\q\q\q\q\q\q
\end{eqnarray}
Now $ Prob_0 (Q_3=1)=P[I]$ is just the probability with no measurement made at $\tau$, and $\delta P(Q_3=1)$
is the  change brought about  by the disturbance produced at $t=\tau$. 
We note that $\delta P(Q_3=1)$ vanishes when $\t$ or 
$T-\t$ equals $k\pi/2$, $k=0,1,2...$ \cite{FOOT2}, and plot its as a function of $\tau$ and $T$, $\tau \le T$ in Fig.3.
In the light-coloured regions, behaviour of the ensemble can be described as  {\it "quantum stochastic"}.
A measurement, added at $t=\tau$ changes the ensemble in Fig. 2b into the one shown in Fig. 2c, 
the number of real paths increases to four, and odds for arriving in the final states at $t=T$, are clearly not what they 
were before.
On the horizontal and diagonal dark lines, the system is {\it "classical stochastic"}. It has only two non-interfering paths \cite{CONS}, leading to different final destinations [see insets a) and c) in Fig. 3].
There is no interference to destroy and, as in classical statistics, we can monitor the system's progress, without disturbing it,
Finally, at the intersection of any two lines, we may call the system's behaviour {\it "classical deterministic"}.
There a single path  [see  inset b) in Fig.3], which leads to a unique final state,
and is travelled every time the experiment is repeated, regardless of whether the measurement at $t=\tau$ is made, or not. 
Note that this classification refers to the present choice of measurements, and choosing a different measured operator, or a different
initial state, would result in a picture, different from the one shown in Fig. 3.
\newline
In general, we note that there can be no pre-determined values (or average values) of $Q_3$, independent of what being done 
at $\t$. Rather, we must conclude that different sets of measurements may "fabricate" completely different statistical ensembles from the same quantum system \cite{DS2}.
\section {No "pre-existing"  path probabilities}
Next we expand on the last statement of the previous Section. 
Let us assume (incorrectly) that there are probabilities to have classical-like pre-determined values 
of $Q_i=\pm1$ at $t_i$, possibly depending on some unknown random classical parameter $\lambda$, [as in  \cite{Bell} we will allow  multiples $\lm=(\lm_1,\lm_2...\lm_M$), in which case $d\lambda$ will mean  $d\lambda_1d\lm_2...d\lm_M$]. With $\lm$ distributed according to some $w(\lm)\ge 0$, $\int w(\lm) d\lm=1$,
 we can evaluate the probabilities for  the sequences of outcomes $Q^{i_3}\gets Q^{i_2}\gets Q^1=1$, 
\begin{eqnarray} \label{2.1}
p[1] =\int p_3(1|1,1,\lm)p_2(1|1,\lm)w(\lm)d\lm,\q\q\n
p[2] =\int p_3(1|-1,1,\lm)p_2(-1|1,\lm)w(\lm)d\lm,\n
p[3] =\int p_3(-1|1,1,\lm)p_2(1|1,\lm)w(\lm)d\lm,\q\n
p[4] =\int p_3(-1|-1,1\lm)p_2(-1,1|\lm)w(\lm)d\lm.
\end{eqnarray}
In Eq.(\ref{2.1}) 
$p_2(Q_2|Q_1=1,\lm)$ 
stands for the probability to have an outcome $Q_2$, given 
a previous outcome $Q_1$, and $p_3(Q_3|Q_2,Q_1=1,\lm)$ yields the odds 
for having a value $Q-3$, given the previous values of $Q_2$ and $Q_1$. 
(Recall that $Q_1$ is always $1$, since the system is prepared in $|+\ra$.)
{\color{black} We expect to have no access to the actual value(s) of the "hidden variable(s)" $\lm$.
We assume, however, that each time the system is set to evolve from its initial state particular path probabilities
$p_3(Q_3|Q_2,Q_1,\lm)p_2(Q_2|Q_1,\lm)$
 exist, 
even if no measurements are made.}
\newline
For our assumption to be correct, we need to demonstrate that the classical probabilities (small $p$'s in Eqs.(\ref{2.1})  are the same as the correct quantum results (capital $P$'s) of the previous Section.
Firstly, we must have
\begin{eqnarray} \label{2.2}
p[i] =P[i], \q i=1,2,3,4.
\end{eqnarray}
Secondly, 
summing the $p[i]$'s over the outcomes at $t=\tau$
we should obtain the probabilities $P(I)$ and $P(II)$ in Eqs.(\ref{0.4}), i.e.,  
\begin{eqnarray} \label{2.3}
Prob(Q_3=1)=P[I]=p[1]+p[2],\q \n
 Prob(Q_3=-1)=P[II]=p[3]+p[4], 
\end{eqnarray}
and, similarly, 
\begin{eqnarray} \label{2.4}
Prob(Q_2=1)=P'[I]=p[1]+p[3],\q \n
 Prob(Q_2=-1)=P'[II]=p[2]+p[4], \q
\end{eqnarray}
However, 
 Eq.(\ref{0.7}) states that, in general,  $P[1]+P[2]\ne P(I)$, so that
 Eqs.(\ref{2.2}) and (\ref{2.3}) cannot always hold.
\newline
This, in turn, demonstrates, that the path probabilities cannot "pre-exist" a set of consecutive
measurements, just as the result of an individual measurement cannot pre-exist the measurement
\cite{Merm}. Different measurements may produce statistical ensembles with distributions as different
as the distributions of heads and tails for differently skewed coins. 
This will happen whenever an additional earlier measurement destroys interference between virtual 
paths leading to later outcomes. 
\section{The negative probability test}
We could look for other proofs of the same point, e.g., by following Feynman's example, described in \cite{Feyn}.
{\color{black} We will not rely on a particular type of quasi-probabilities, as was done, for example, in \cite{Hall}, but rather 
assume that the classical-like path probabilities $p[i]$, similar to those in (\ref{2.1}), can somehow be defined.
 We will then look for the  values they must take in order to reproduce 
the correct quantum mechanical results.}
With the help of Eqs.(\ref{2.1})  it is easy to express the average values of the products, $\la Q_i Q_j\ra_{cl}$,
in terms of the $p[i]$'s,
\begin{eqnarray} \label{1.2}
\la Q_1Q_2\ra_{cl}=p[1]-p[2]+p[3]-p[4],\n
\la Q_1Q_3\ra_{cl}=p[1]+p[2]-p[3]-p[4],\n
\la Q_2Q_3\ra_{cl}=p[1]-p[2]-p[3]+p[4].
\end{eqnarray}
Using the path probabilities in Eqs. (\ref{0.5})-(\ref{0.7})  yields the correct quantum 
value for the same quantities
\begin{eqnarray} \label{1.3}
\la Q_1Q_2\ra \equiv P'[I]-P'[II]=\cos(2\t)\equiv \alpha,
\q\q\q\q\q
\n
\la Q_1Q_3\ra \equiv P[I]-P[II]=\cos(2T)\equiv \beta,\q\q\q\q\q
\n
\la Q_2Q_3\ra \equiv P[1]-P[2]-P[3]+P[4]=
\q\q\q\q\q\n
\cos(2(T-\t))\equiv \gamma,\q\q\q\q\q
\end{eqnarray}
If our assumption 
 is correct, results (\ref{1.2}) and (\ref{1.3}) will agree.
Equating $\la Q_iQ_j\ra_{cl}=\la Q_iQ_j\ra$, and adding a condition 
\begin{eqnarray} \label{1.3a}
p[1]+p[2]+p[3]+p[4]=1,
\end{eqnarray}
yields four linear equations the probabilities $p[i]$ must satisfy.
Their solutions are 
\begin{eqnarray} \label{1.4}
p[1]=(\alpha+\beta+\gamma +1)/4,\q\n
p[2]=(-\alpha+\beta-\gamma +1)/4,\n
p[3]=(\alpha-\beta-\gamma +1)/4,\q\n
p[4]=(-\alpha-\beta+\gamma +1)/4.
\end{eqnarray}
Our assumption 
 will be proven wrong, if at least one of the $p[i]$'s turned out 
to be negative. 
Thus, we evaluate
\begin{eqnarray} \label{1.5}
\delta p(\t,T) \equiv  \sum_{i=1}^4 |p[i]|-1, 
\end{eqnarray}
which is zero if, and only if, all $p[i]$'s are non-negative, and map it on the $(\t,T)$ plane in Fig. 4. 
As in Fig.3, $\delta p(\t,T)\ne 0$ in the light-coloured regions, and vanishes on the horizontal and diagonal 
lines where, as we already  know from Sect.III, the classical-like probabilities can be defined.
We note that this negative probability test is also passed also on the vertical lines
$T=k\pi/2$, $k=1,2,...$. An inspection of Eqs.(\ref{1.4}) shows that for any $\tau$, and $T=(2k+1/2)\pi$, or $T=k\pi$
 there exists a suitable classical ensemble. Such ensembles, with only two paths leading to the same destination,
are shown 
 in the insets if Fig. 4.  Just because such ensembles can be found in principle,
 does not, of course, mean that they correspond to what actually happens.
 To warn the reader about the misrepresentation, we crossed the insets in Fig. 4 with red lines.
\newline
Clearly, the appearance of negative "probabilities" is a sufficient, yet not necessary
condition for the classical-like reasoning, based on Eqs.(\ref{1.2}), to fail.
Such is the price of relying on average values, instead of the probability distributions,
which contain full information about a statistical ensemble.
Relying on the properties of sums of averages, rather than on the averages themselves,  
would be an even less precise tool, as we will discuss next.
\begin{figure}[h]
\includegraphics[angle=0,width=8.5cm, height= 7cm]{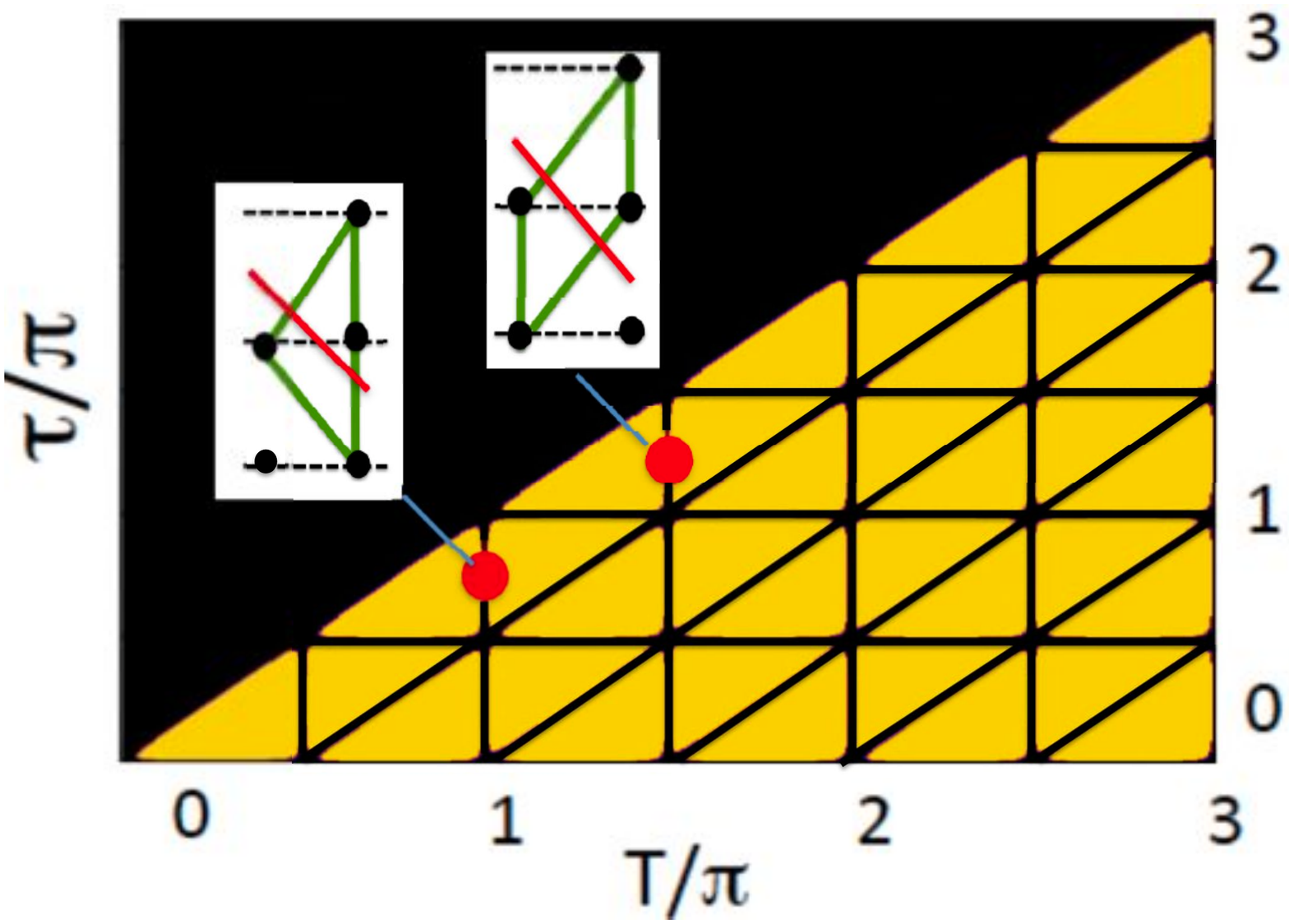}
\caption {Detection of "signalling in time" by the negative probability test.
In the light coloured area $\delta p(\t,T) \ne 0$ in Eq.(\ref{1.5}), while on the dark coloured
lines one can find classical path probabilities, consistent with 
the quantum averages (\ref{1.3}).  The ensembles the test finds on the 
vertical lines (see the insets) are, however, incorrect.
The true ensembles for these values of $\t$ and $T$ are shown in Fig.2c. }
\label{fig:FIG1}
\end{figure}
\section{The Leggett-Garg inequalities}
Alternatively, we might note that the existence of the classical-like non-negative path probabilities (\ref{2.1})
imposes certain restrictions on the sums of the averages (\ref{1.2})
Following  \cite{LGI} one notes that in all of the four possible sequences,  
$Q_3\gets Q_2\gets Q_1$
the sum of products $L=Q_1Q_2 +Q_1Q_3+Q_2Q_3$ equals $3$, if all the $Q$'s have the 
same sign, and takes the value of $-1$ otherwise. It is readily seen that if the sequences occur with the probabilities 
in Eq.(\ref{2.1}), also the sum of the averages (\ref{1.2}) cannot be smaller that $-1$,
\begin{eqnarray} \label{4.1}
\la L\ra \equiv \la Q_1Q_2 \ra +\la Q_1Q_3\ra +\la Q_2Q_3\ra \ge -1,
\end{eqnarray}
since a chance to add the $1\gets 1\gets 1$, sequence would only increase the value of $L$.
The LGI test consist in inserting the correct quantum values 
(\ref{1.3}) into (\ref{4.1}) and looking for the values of $\t$ and $T$, such that the inequality does not hold.  
Thus, we will look for those values of $\t$ and $T$, for which the sum
\begin{eqnarray} \label{4.1}
\delta L(\t,T) \equiv \la L\ra+1=\alpha+\beta+\gamma +1,
\end{eqnarray}
where $\alpha$, $\beta$ and $\gamma$ are defined in Eq.(\ref{1.3}), is negative.
A condition $\delta L < 0$ should, therefore, signal the impossibility of assigning 
meaningful path probabilities $p[i]$
in (\ref{1.2}), in the same way as the appearance 
of "negative probabilities", discussed in the previous Section. 
This is, however, a less direct 
approach, and we ask whether it is as efficient as the tests of the previous two Sections.
\newline
We already know that the LGI would be satisfied on the network of lines in Fig.4, 
since a suitable classical ensemble does exist on its the horizontal and diagonal lines,
while on the vertical lines it  can be found at laest in principle. 
Indeed, these lines divide the ($\tau$, $T$)-plane into the segments 
inside which the LGI is either violated (light colour), or satisfied (black), as shown in Fig.5. 
It is readily seen that the LGI leaves much of the ($\t,T$)-plane black, being
a much less sensitive indicator of "signalling in time", than the negative probability test of Sect.V.
Such is the price of relying on the sums rules, satisfied by the averages, rather than on the averages themselves.
\begin{figure}[h]
\includegraphics[angle=0,width=9cm, height= 8cm]{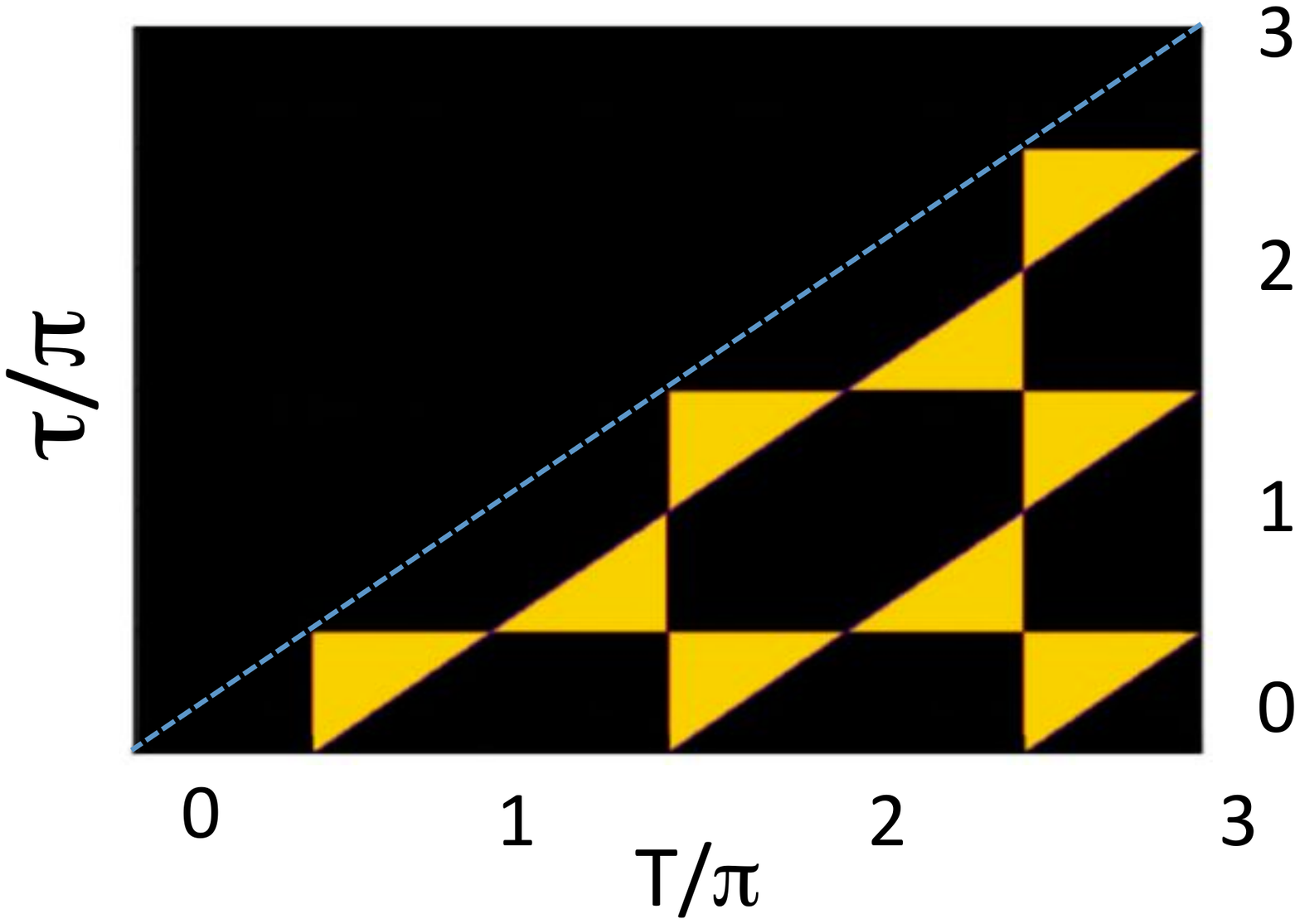}
\caption {Detection of "signalling in time" by the Leggett-Garg inequality.)
In the light coloured area quantum mechanical averages (\ref{1.3}) violate 
the Leggett-Garg inequality (\ref{4.1}), while in the dark coloured areas the inequality holds.}
\label{fig:FIG1}
\label{fig:FIG1}
\end{figure}
\section{Conclusions and discussion}
In summary, a sequence of quantum measurements, made on an elementary quantum system, 
can 
be described in terms of real observable paths, constructed from the virtual ones. 
The outcomes of previous observations can influence the results of the measurements
yet to be made, provided the earlier measurements create new real scenarious, by destroying 
interference, otherwise existent between the virtual paths.
\newline
A simple illustration of such a description may be provided by a qubit, completing  its Rabi cycle by $t=T=\pi$. 
In this case, two interfering paths, $\{3\}$ and $\{4\}$ in Fig.1, have
amplitudes of the same magnitude, but of opposite sign, $\pm i\sin \t\cos \t $.
Destructive interference  prevents  the system 
from reaching the state  $|-\ra$. An additional measurement at $t=\t$ makes both virtual paths real as 
shown if Fig. 2c, 
and, at $t=T$, the qubit is found in $|-\ra$ with a probability $\sin^2( \t)/2$. An earlier measurement
at $t=\tau$ clearly affect the outcomes at $T$, or, if one prefers the language of \cite{Koff}, 
"signalling in time" occurs.  
\newline
We also considered two other approaches, based on evaluation of two-times averages of the qubit's variable.
One approach assumes that the real scenarios (paths) exist at all times, and are not created by the measuring device(s). 
It fails,  since meaningful path probabilities cannot, in general, be found where destruction of interference between virtual paths 
is known to take place. One exception are the vertical lines in Fig.4, where this "negative probability test"
errs by finding a spurious ensemble, consistent with the quantum mechanical averages (\ref{1.3}), but misrepresenting
the actual situation.
\newline
The second method, based on the Leggett-Garg inequalities, tests a sum rule, which the averages 
should satisfy in the absence of "signalling". As suggested in \cite{Koff}, the LGI provide a sufficient, but not necessary 
condition, and detects the quantum behaviour in far fewer cases than the negative probability test, as shown in Fig. 5.
Perhaps, one reason for the popularity of the approach is the LGI's formal similarity to the celebrated Bell's inequality \cite{Bell}. We find, however, little advantage in using the analogy, and advocate much simpler elementary methods, serving the same purpose.
\section{Acknowledgements}
Financial support of
MINECO and the European Regional Development Fund FEDER, through the grant
FIS2015-67161-P (MINECO/FEDER,UE) and the Basque Government Grant No IT986-16
is acknowledged by DS.

 

\begin{thebibliography}{10}
 \bibitem{nat}  Knee, G.C., et al.
A strict experimental test of macroscopic realism in a superconducting flux qubit,
{\it  Nat. Comm.},   [7:13253], DOI: 10.1038/ncomms13253, (2016).
  \bibitem{LGI}  Leggett, A.J. \&  A. Garg, A., 
 Quantum Mechanics versus Macroscopic Realism: Is the Flux There when Nobody Looks?
    {\it Phys. Rev. Lett.} {\bf 54}, 857 (1985).
  \bibitem{Koff} Kofler. J. \& Brukner, C., 
 Condition for macroscopic realism beyond the Leggett-Garg inequalities.
    {\it Phys. Rev. A}, {\bf 87}, 052115 (2013).  
  \bibitem{Bell} Bell, J.S.,
   On the Einstein Podolsky Rosen paradox,
  {\it Physica} {\bf 1}, 195 (1964). 
  \bibitem{noclo} Wooters, W. \& Zurek, W.,
  A single qubit cannot be cloned.
   {\it Nature} {\bf 299}, 802 (1982).
 \bibitem{Feyn} Feynman, R.P.,
 Simulating physics with computers.
   {\it Int. J. Theor. Phys.}  {\bf 21}, 467 (1982).
 \bibitem{Whit} Whitaker, A., 
  Richard Feynman and Bell's theorem.
   {\it Int. J. Theor. Phys.}  {\bf 21}, 467 (1982).
  \bibitem{Hall}Haliwell, J.J.,
   Leggett-Garg inequalities and no-signaling in time: A quasiprobability approach.
 {\it  Phys. Rev. A} {\bf 93}, 022123 (2016). 
   \bibitem{Em1} Emary, C.,
 Ambiguous measurements, signalling, and violations of Leggett-Garg inequalities.
   {\it  Phys. Rev. A} {\bf 96}, 042102 (2017). 
\bibitem{Foot} We note that "signalling in time" may be a rather fancy description of what happens.
A tennis player, returning the ball the his/her partner may be said to "have sent a signal 
forward in time". However, "hitting the ball back" would usually do.
   \bibitem{Em2}  Emary,  D., N. Lambert, F. Nori,
  Leggett-Garg inequalities.
   {\it  Rep. Prog. Phys.} {\bf 77}, 016001 (2014).
 \bibitem{Feynl} R.P. Feynman, A.R. Hibbs,, 
  {\it Quantum mechanics and path Integrals}
(McGrawHill, New York, 1965).
 \bibitem{DS1} Sokolovski, D.,
  Quantum measurements, stochastic networks,
the uncertainty principle, and the not so strange
Òweak valuesÓ.
{\it Mathematica} {\bf 4}, 56 (2016), doi:10.3390/math4030056 .   
 \bibitem{DS2} Sokolovski, D.,
 Path probabilities for consecutive measurements, and certain Òquantum paradoxesÓ.
  {\it  Ann. Phys,} {\bf 397}, 474 (2018).
    \bibitem{FOOT2}
  An analogy with the Young's interference experiment may be helpful. 
Treating the states at $\t$ and $T$  as two "slits", and two "positions on the screen", respectively, we note that the first case corresponds to one of the slits blocked . Now the paths leading to final positions can be identified without destroying the pattern on the screen. The case $\sin(2(T-\t)))=2\sin(T-\t)\cos(T-\t)=0$
corresponds to both slits being open, but just one path  [$\{1\}$ and $\{4\}$, if  $\sin(T-\t)=0$,
or $\{2\}$ and $\{3\}$, if  $\cos(T-\t)=0$] connecting each slit with each final position.
Again, the "which way?" question can be answered, since there is no interference to destroy. 
 \bibitem{CONS} Note that these non-interfering paths can be considered "consistent histories"
 in the consistent histories approach, 
 Griffiths, R.B., Consistent quantum measurements.  {\it Stud. Hist. Phil. Mod. Phys.} {\bf 52} 188, (2015).
 \bibitem{Merm} Mermin, N.D.,  Hidden variables and the two theorems of John Bell.  {\it Rev. Mod. Phys.} \textbf{65}, 803 (1993).
%
\end{thebibliography}
\end{document}